\documentclass[submission, Proceedings]{SciPost}

\hypersetup{
    colorlinks,
    linkcolor={red!50!black},
    citecolor={blue!50!black},
    urlcolor={blue!80!black}
}
\usepackage[square,sort,comma,numbers]{natbib}

\usepackage{afterpage}
\usepackage{subcaption}
\usepackage[bitstream-charter]{mathdesign}
\let\OLDthebibliography\thebibliography
\renewcommand\thebibliography[1]{
  \OLDthebibliography{#1}
  \setlength{\parskip}{0pt}
  \setlength{\itemsep}{0pt plus 0.3ex}
}
\DeclareSymbolFont{usualmathcal}{OMS}{cmsy}{m}{n}
\DeclareSymbolFontAlphabet{\mathcal}{usualmathcal}

\begin{document}

\begin{center}{\Large \textbf{
Azimuthal di-jet correlations with parton branching TMD distributions\\
}}\end{center}

\begin{center}
A.~Bermudez Martinez$^a$ and 
F.~Hautmann$^{b, c, d}$
\end{center}

\begin{center}
{$^a$Deutsches Elektronen-Synchrotron DESY, 
Notkestr. 85, 22607 Hamburg, Germany}
\\
{$^b$CERN,  Theoretical Physics Department, 
CH 1211 Geneva 23, Switzerland 
}
\\
{$^c$Elementaire Deeltjes Fysica, Universiteit 
Antwerpen, B 2020 Antwerpen, Belgium
}
\\
{$^d$RAL, Chilton OX11 0QX and University of 
Oxford, Oxford   OX1 3PU, UK  
}
\\
\end{center}


\definecolor{palegray}{gray}{0.95}
\begin{center}
  \begin{tabular}{rr}
  &
  \begin{minipage}{1.0\textwidth}
    \begin{center}
    {\it Presented at DIS2022: XXIX International Workshop on Deep-Inelastic Scattering and Related Subjects, Santiago de Compostela, Spain, May 2-6 2022} \\
    \end{center}
  \end{minipage}
\end{tabular}
\end{center}

\vspace*{-8.5cm}
\begin{flushleft}
CERN-TH-2022-137  
\;\;\;\; DESY-22-135  \\
\end{flushleft}
\vspace*{+7.0cm}

\section*{Abstract}
{\bf
The parton branching formulation of TMD evolution has recently been used to make  predictions for 
jet observables at the Large Hadron Collider (LHC), including perturbative 
matching at next-to-leading order  (NLO). This 
contribution presents results for the azimuthal $\Delta \phi$ correlations in events 
with di-jets at large transverse momentum. It focuses on the back-to-back region of large $\Delta \phi$  
and discusses prospects for detailed studies of QCD dynamics in this region at the LHC. 
}



\section{Introduction}
\label{sec:intro}

Azimuthal correlations between two jets have been measured at the LHC 
by the ATLAS and CMS 
collaborations~\cite{daCosta:2011ni,Khachatryan:2011zj,Khachatryan:2016hkr,CMS:2017cfb,CMS:2019joc}. A detailed understanding of 
these correlations is important for studies of the 
Quantum Chromodynamics (QCD) sector of the Standard Model (SM) and 
 searches for Beyond-the-SM (BSM) scenarios with di-jet signatures. 

At leading order (LO) in the strong coupling $\alpha_s$, two jets are produced   
 back-to-back, i.e., with azimuthal angle $\Delta \phi = \pi$. Deviations from this 
configuration  measure higher-order QCD radiation. In the region 
near $\Delta \phi = \pi$  this is primarily soft gluon radiation, while in the region of 
small $\Delta \phi $ it is primarily hard QCD radiation. 
 Since initial-state parton radiation moves the jets away from the 
 $\Delta \phi = \pi$   region, it is relevant to investigate the 
 influence of transverse momentum recoils in the 
 QCD showers~\cite{Hautmann:2008vd,Jung:2010si,Dooling:2012uw}, 
 taken into account via 
 transverse momentum dependent  (TMD)~\cite{Angeles-Martinez:2015sea}  
parton  distributions, 
 on the description of the $\Delta \phi$   measurements. 
 
In this  article we discuss this by using the Parton Branching (PB) 
approach~\cite{Hautmann:2017xtx,Hautmann:2017fcj} to 
TMD    distributions.   
 This approach has successfully been 
used at next-to-leading order (NLO) 
to extract TMD parton distributions~\cite{BermudezMartinez:2018fsv} 
from precision deep-inelastic data~\cite{disdata15} using    
xFitter~\cite{HERAFitter,xFitterDevelopersTeam:2022koz}
(results are available from the 
repository~\cite{Abdulov:2021ivr,Hautmann:2014kza}). 
It has also been successfully used 
 to make predictions for Drell-Yan transverse momentum spectra 
 both at the LHC~\cite{BermudezMartinez:2019anj} 
 and in lower energy experiments~\cite{BermudezMartinez:2020tys}. 
We here apply this approach to di-jet production,  presenting results 
from the work in Ref.~\cite{Abdulhamid:2021xtt}. 
We compute predictions 
for di-jet azimuthal correlations, using the PB TMD evolution matched with 
NLO perturbative matrix elements.

In the region near the back-to-back configuration 
the QCD Sudakov process depends on the soft 
function~\cite{Collins:1999dz}. 
Unlike the case of Drell-Yan di-lepton production, 
 factorization breaking 
effects~\cite{Rogers:2010dm,Vogelsang:2007jk,Collins:2007nk} can arise  in the case of di-jets 
 due to long-timescales soft-gluon correlations 
between initial and final states. 
We examine the possibility of investigating these 
effects with high transverse momentum jets at the 
LHC. 

 The article is organized as follows. In Sec.~2  we briefly describe the 
 main elements of the 
 PB TMD calculation at NLO. 
 In Sec.~3 we illustrate the results for  di-jet azimuthal distributions,   and
 compare them with 
 LHC 
experimental measurements. 
   Conclusions are given in Sec.4.

\section{NLO matching with PB TMD}
\label{sec-method}

The PB approach~\cite{Hautmann:2017xtx,Hautmann:2017fcj} provides 
evolution equations for TMD distributions in terms of Sudakov form factors 
and splitting probabilities, and a corresponding TMD parton shower in a backward 
evolution scheme.  
PB TMD distributions and parton showers are implemented in the 
Monte Carlo event generator {\sc Cascade}3~\cite{Baranov:2021uol}. 

A method to match  TMD evolution with NLO perturbative matrix elements 
has been developed for the case of the Drell-Yan process in 
Refs.~\cite{BermudezMartinez:2019anj,BermudezMartinez:2020tys} 
using the framework of {\sc MadGraph5\_aMC@NLO}~\cite{Alwall:2014hca}. 
We next apply this method to the case 
of the jet production process~\cite{Abdulhamid:2021xtt}, matching 
 PB TMD distributions and parton showers 
with  di-jet NLO matrix elements from 
 {\sc MadGraph5\_aMC@NLO}. 
Further details on the NLO matching method with PB TMD are given in 
Ref.~\cite{Yang:2022qgk},  
where a comparison of 
  MCatNLO+{\sc Cascade}3 \cite{Baranov:2021uol} 
and 
MCatNLO+{\sc Herwig}6 \cite{Corcella:2002jc} 
matching 
is performed. 

\begin{figure}
\centering
\includegraphics[width=10cm]{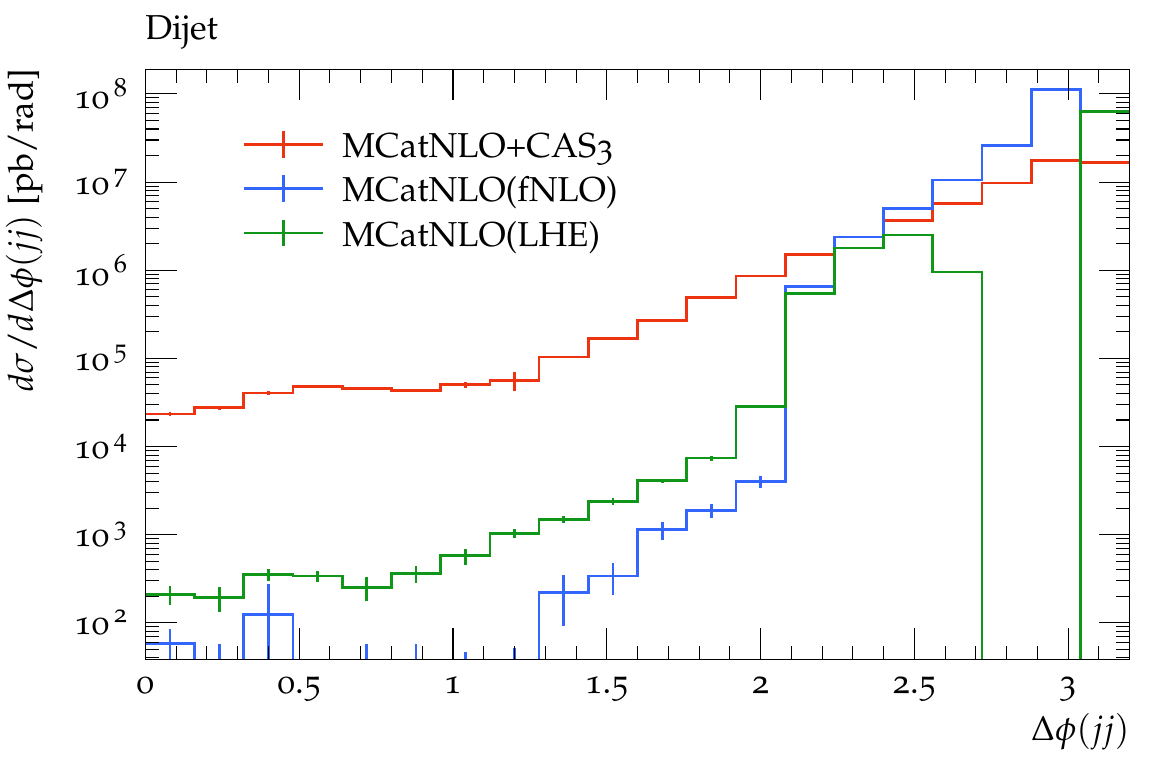}
\caption{Azimuthal $\Delta \phi$ distributions obtained from  
the fixed NLO calculation (MCatNLO(fNLO)), 
the (unphysical) LHE level (MCatNLO(LHE)), 
and after inclusion of 
PB  TMDs (MCatNLO+CAS3)~\protect\cite{Abdulhamid:2021xtt}.}
\label{fig:fixed-lhe-pb}
\end{figure}

Fig.~\ref{fig:fixed-lhe-pb} illustrates the result of applying the matching method. 
It 
 shows the differential distribution in the azimuthal angle 
 $\Delta \phi$  between the two leading jets 
 as obtained from the  
   calculations at fixed NLO (blue curve), at the (unphysical) level 
   including the  subtraction terms from the  matching  
 (LHE level, green curve), and after inclusion of  PB TMDs (red curve).
We observe the rising cross section of the fixed NLO calculation towards 
large  $\Delta \phi$ (corresponding to the 
divergent behavior of the NLO calculation in 
the back-to-back configuration), the decay  towards large $\Delta \phi$ 
once  the subtraction terms are included, and the  smooth prediction 
once the TMD distributions and showers are included.

\section{Di-jet azimuthal distributions}
\label{sec-results}

We now use the method of the previous section to compute 
NLO-matched PB TMD predictions for di-jet distributions in the 
phase space of the 
CMS measurements~\cite{CMS:2017cfb,CMS:2019joc}. 

We consider selection cuts for 
leading jets with 
transverse  momentum $ p_T > 200 $ GeV 
and $ p_T > 1000 $ GeV.  
With this event selection one is able to explore 
TMD dynamical effects over a broad range both in the 
transverse momentum of the TMD distribution, 
set by the $p_T$ imbalance between the jets, 
and in its  
evolution scale, set by the 
hard scale of the event, e.g.~the leading jet $p_T$. 
In particular, in a neighborhood  of order  0.1 rad  
from $\Delta \phi = \pi$, the $p_T$ imbalance 
ranges from a few ten GeV for the highest $p_T$ jets down to 
few GeV. At large $p_T$ imbalance, the 
evolution of the transverse 
momentum is dominated by perturbative contributions to the 
evolution kernels and can 
be explored through directly measurable  jets, while  at lower 
$p_T$ imbalance  both perturbative and 
non-perturbative components can be investigated. 

\begin{figure}
\centering
\includegraphics[width=0.45\textwidth]{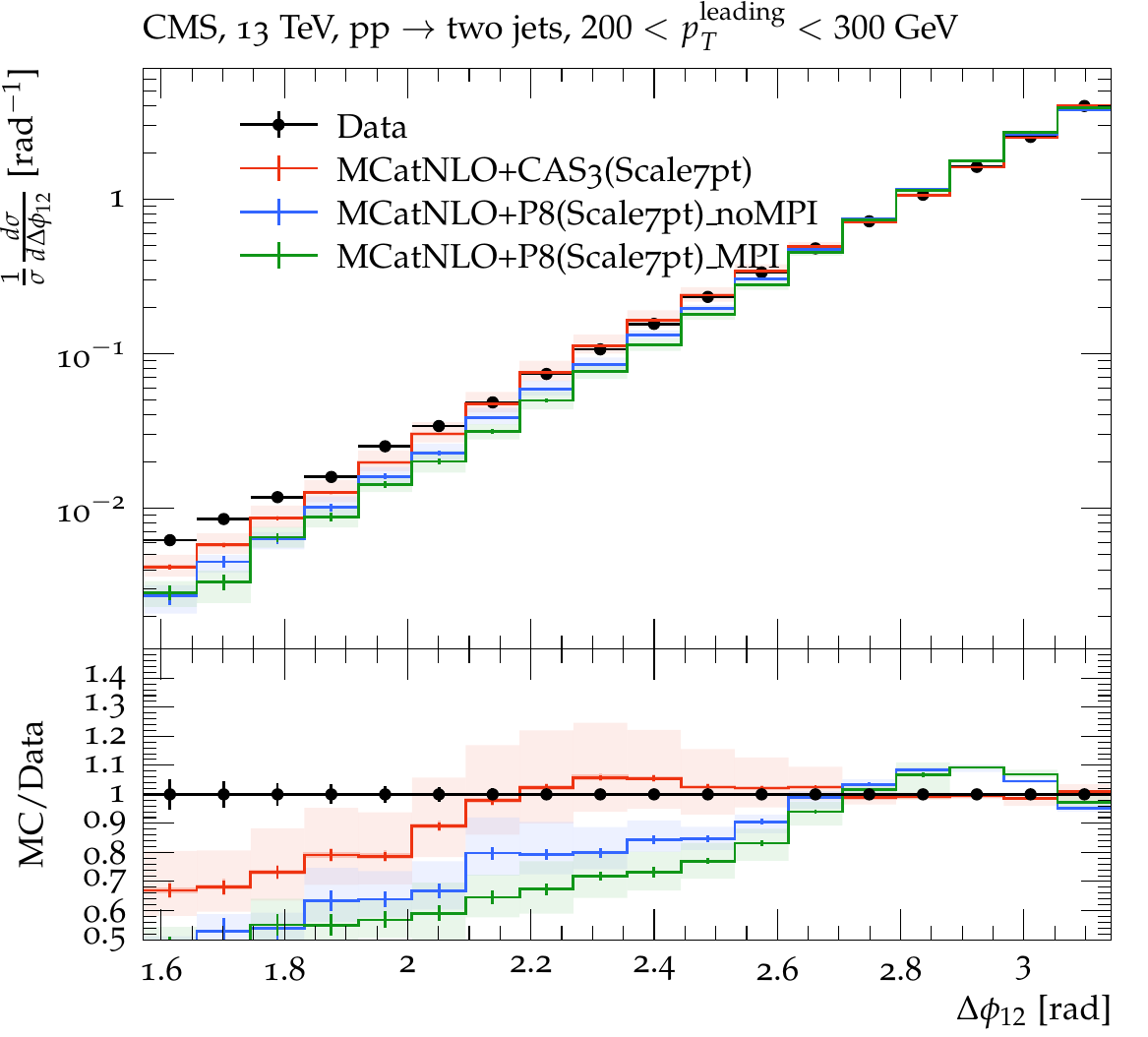} 
\includegraphics[width=0.45\textwidth]{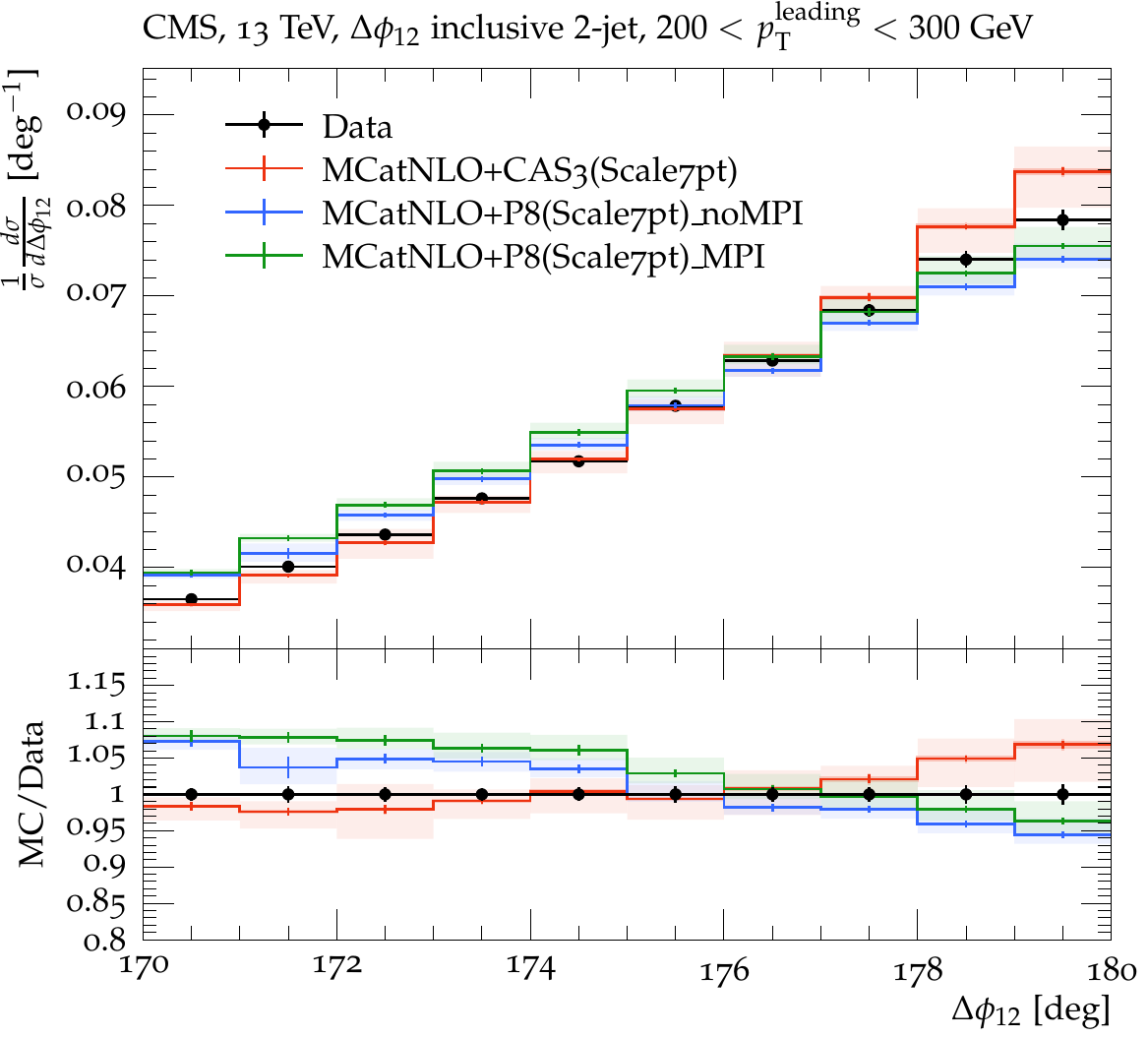} 
\caption{Azimuthal correlation   over a wide $\Delta \phi$ range  (left) and 
 in the back-to-back region (right)~\protect\cite{Abdulhamid:2021xtt}. 
 CMS data~\protect\cite{CMS:2017cfb,CMS:2019joc} 
 are  compared with 
 results from MCatNLO+{\sc Pythia}8  and 
  MCatNLO+CAS3.}  
\label{fig:cas-p8}
\end{figure}

In Fig.~\ref{fig:cas-p8} we report the 
NLO-matched PB TMD 
results (labelled 
MCatNLO+CAS3), together  with 
 CMS data~\cite{CMS:2017cfb,CMS:2019joc} and collinear shower  
calculations  from 
 MCatNLO+{\sc Pythia}8 \cite{Sjostrand:2014zea}. 
 The shape of the $\Delta \phi$ distribution is different between 
the TMD and collinear shower calculations, 
emphasizing the relevance of the detailed dynamics of 
QCD shower evolution. 
The uncertainty bands on the MCatNLO+CAS3 predictions are obtained 
from scale variations and TMD uncertainties~\cite{Abdulhamid:2021xtt}. 
The uncertainty bands on the MCatNLO+{\sc Pythia}8 predictions 
are obtained from scale and associated shower variations according to the 
method of~\cite{Mrenna:2016sih} together with the 
guidelines of~\cite{Gellersen:2020tdj}. 
For the MCatNLO+{\sc Pythia}8 
calculation the effect of multi-parton interactions (MPI) is also 
shown, using the parameters of the 
tune CUETP8M1~\cite{Khachatryan:2015pea}. 
For leading jet $p_T > 200$ GeV, 
the MPI effect is not large.  

 MCatNLO+CAS3 describes the measurements well at large and 
 intermediate $\Delta \phi$. In the decorrelated 
 region  at low $\Delta \phi$ a deficit is observed. 
 This  
  is due to missing higher-order contributions from multiple 
 QCD emissions beyond NLO. To take these contributions 
 into account, one needs to go beyond the framework 
 of the present calculation, for example by employing TMD 
 multi-jet merging 
 techniques~\cite{Martinez:2021chk,Martinez:2022wrf}.  

\begin{figure}
\centering
\includegraphics[width=0.45\textwidth]{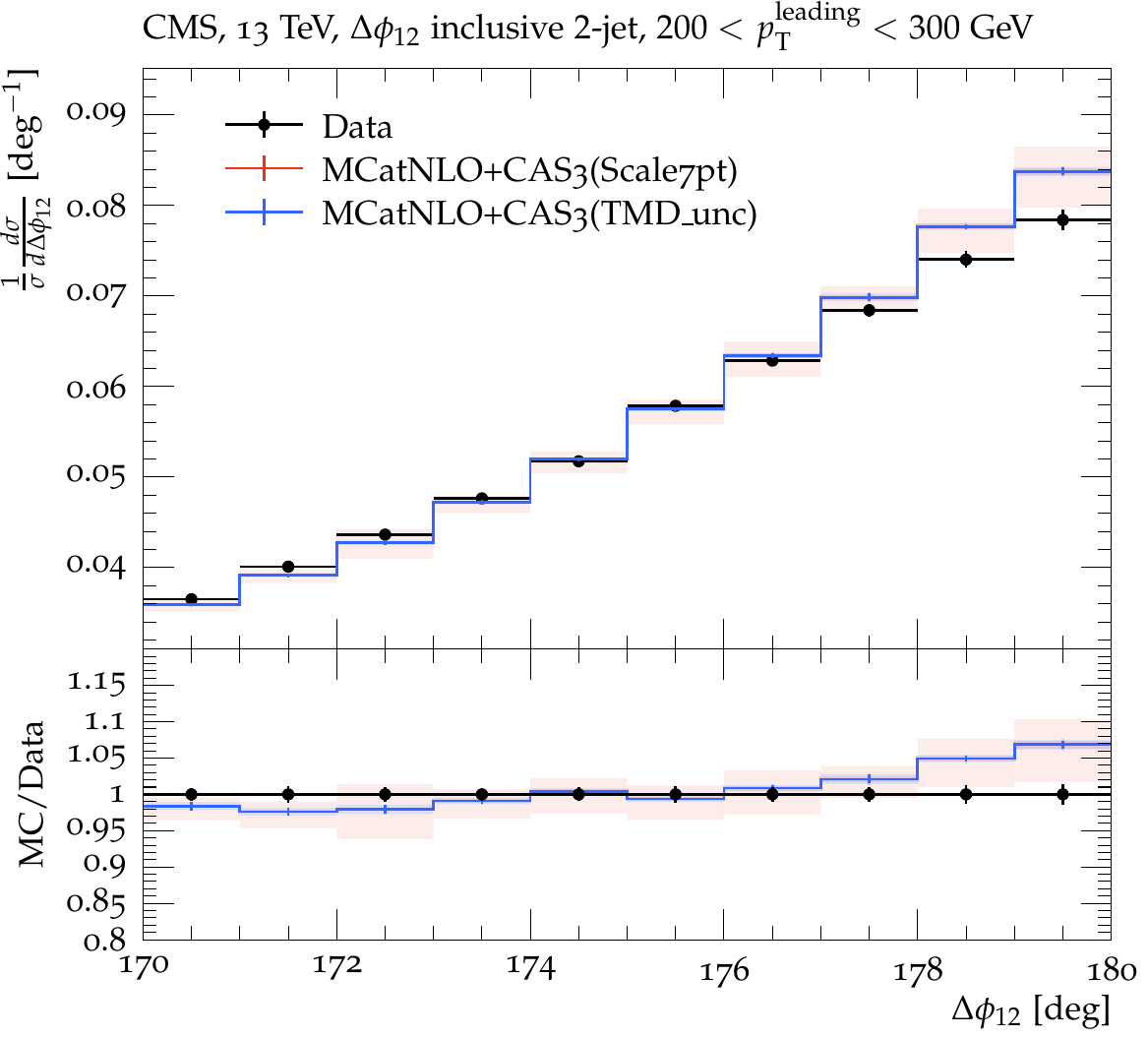} 
\includegraphics[width=0.45\textwidth]{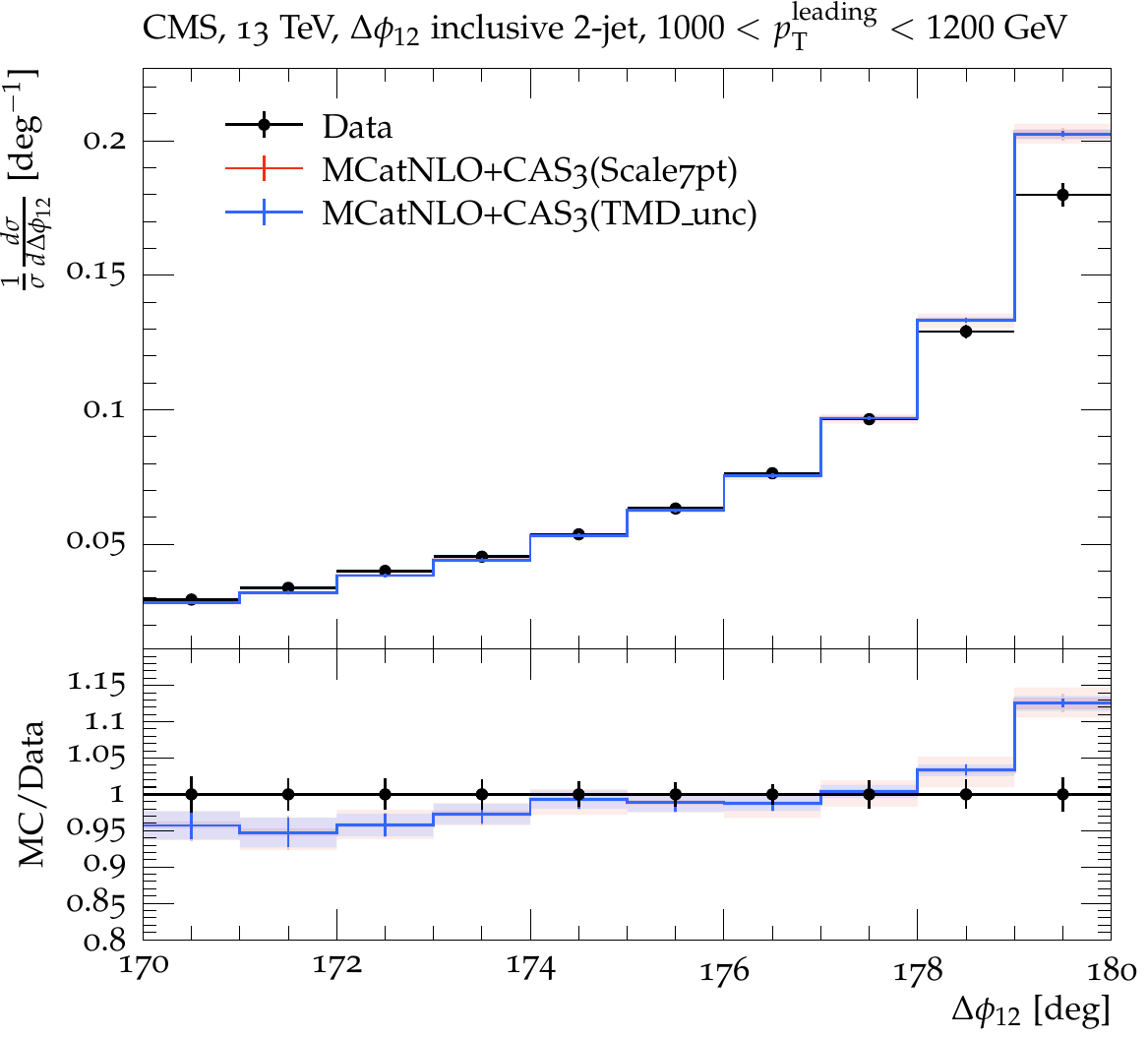} 
\caption{Azimuthal correlation   in the back-to-back region for leading jet 
$p_T > 200 $ GeV (left) and $p_T > 1000 $ GeV (right) as measured by 
CMS~\protect\cite{CMS:2019joc} compared with predictions from 
MCatNLO+CAS3~\protect\cite{Abdulhamid:2021xtt}. 
Shown are the uncertainties coming from the scale variation 
as well as the uncertainties coming from the TMD.}
\label{fig:largeDeltaPhi}
\end{figure}

In Fig.~\ref{fig:largeDeltaPhi} we focus on the large $\Delta \phi$ 
region of nearly back-to-back jets. 
This region is of special interest,  as 
possible factorization-breaking effects have long been conjectured to 
arise  for back-to-back jets 
 due to soft-gluon interactions between initial and final states.  
We see from  Fig.~\ref{fig:largeDeltaPhi} that the 
measurements are well described 
by the MCatNLO+CAS3 predictions. 
Only in the highest   bin ($\Delta \phi > 179^o$) 
a deviation of about 10\% is observed.   
Detailed phenomenological studies in this region 
are warranted, using fine binning in angle.  

As discussed in Ref.~\cite{Yang:2022qgk}, 
further insight may be gained from the combined 
analysis of $\Delta \phi$  correlations in  di-jet and 
$Z$-boson + jet events. 
At low $p_T$ the boson-jet state 
is more strongly correlated azimuthally 
than the jet-jet state, while for $p_T$ far above the 
electroweak  scale 
the behaviors become more similar. 
This can be connected to features of the partonic 
initial-state and final-state radiation in the boson-jet and jet-jet cases. 
Initial-state 
 and final-state 
 radiation (see the 
 recent studies~\cite{Bouaziz:2022vp,Chien:2022wiq} 
 in the $Z$ + jet process) 
 may give rise to color interferences and 
potential factorization-breaking 
effects~\cite{Rogers:2010dm,Collins:2007nk,Rogers:2013zha}. 
   If so, different  breaking patterns can be expected for  
 strong and weak azimuthal correlations,    
 influencing differently the boson-jet and jet-jet cases. 
 Systematic measurements of 
 di-jet  and $Z$-boson + jet distributions are 
thus  proposed~\cite{Yang:2022qgk}, scanning the 
 phase space  from low transverse momenta $p_T \approx {\cal O}$(100 GeV) 
 to high transverse momenta   $p_T \approx {\cal O}$(1000 GeV). 

\begin{figure}
\centering
\includegraphics[width=0.45\textwidth]{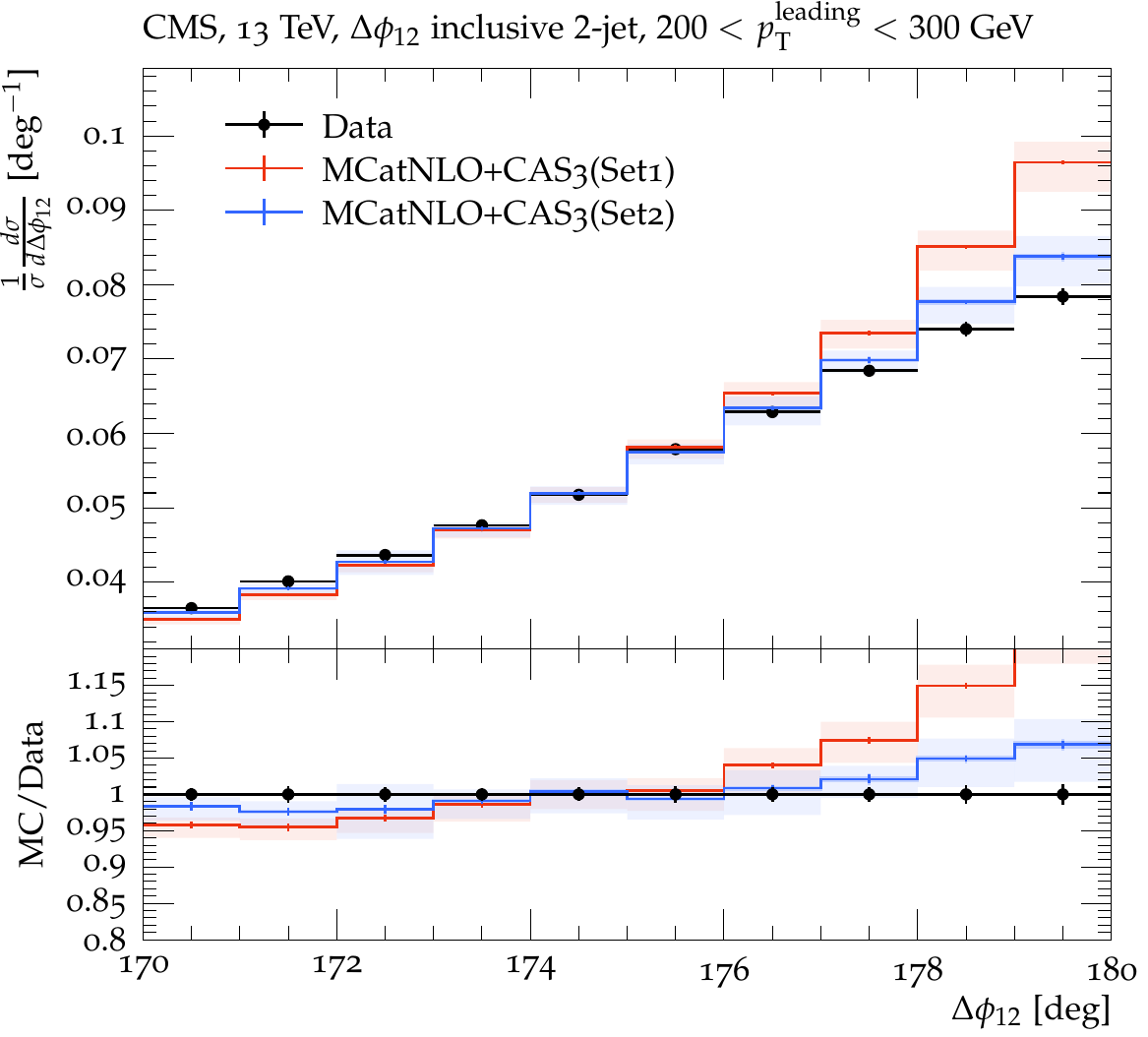} 
\caption{Impact of the 
transverse momentum $q_T$ scale in the running coupling 
at large $\Delta \phi$~\protect\cite{Abdulhamid:2021xtt}. As discussed in the text,  
the $q_T$ scale is used in the result Set 2, not in the result Set 1.}
\label{fig:alphas_qt}
\end{figure}

Fig.~\ref{fig:alphas_qt}    
 illustrates  another aspect of the QCD dynamics in the back-to-back 
region, namely, the sensitivity to 
soft-gluon angular 
ordering~\cite{Bassetto:1983mvz,Dokshitzer:1987nm,Marchesini:1987cf,Catani:1990rr} 
in the TMD evolution~\cite{Hautmann:2017fcj,Hautmann:2019biw}. 
The MCatNLO+CAS3 curves labelled 
 Set1 and Set 2 in Fig.~\ref{fig:alphas_qt}   
 refer to two sets of PB TMD distributions~\cite{BermudezMartinez:2018fsv}
differing  by the  scale in the QCD running coupling: 
Set 2  fulfills the soft-gluon angular ordering conditions by using  
the transverse momentum emitted at each branching 
as a scale for $\alpha_s$, 
while Set 1  uses the branching scale  
as a scale for $\alpha_s$, as in DGLAP ordered evolution.    
We see that the shape of the 
azimuthal correlation  is sensitive to angular ordering effects 
in the back-to-back region. 
Set 2 provides a better description of the measurements in this region.

\section{Conclusion}
\label{sec-conc}

In this article  we have discussed 
 predictions from PB TMD parton showers   
  for  final state observables in jet production at the LHC, 
  focusing on the azimuthal correlations  of jets  
  with large transverse momenta. 

The PB TMD shower  matched with NLO calculations provides a good 
description of 
experimental measurements of  di-jet production at the LHC 
in the correlation region  of high  azimuthal separations 
 $\Delta \phi$ between the jets, down to the region of 
 intermediate  $\Delta \phi$. 
The shape of the $\Delta \phi$ distribution is sensitive to the 
detailed dynamics of the shower evolution. We have studied 
effects of TMD versus collinear shower and of soft-gluon 
angular ordering.  

In the back-to-back region near  $\Delta \phi = \pi$,   
potential factorization breaking contributions can  arise 
due to colored final states. We have discussed that 
these effects can be explored through 
future dedicated measurements 
with large-$p_T$ jets and fine binning in $\Delta \phi$. 

In the decorrelated region of  low $\Delta \phi$, we 
observe a deficit in the predictions 
 due to  missing higher-order 
contributions from multiple QCD emissions beyond NLO. Including multiple 
emissions  requires further methodologies, for instance 
multi-jet merging~\cite{Martinez:2022wrf}, which 
have not yet been applied here. 

Given the 
observed sensitivity of the  $\Delta \phi$ distribution 
to  angular ordering, 
it will   be of interest to include recent developments of TMD branching 
such as the parton distribution fits with 
angular-ordered resolution scale~\cite{Barzani:2022msy}. 
Also, it will be relevant to investigate the role of 
the recently proposed      branching with 
 TMD splitting functions~\cite{Hautmann:2022xuc} on the 
azimuthal asymmetries.

\section*{Acknowledgments}

The results discussed in this contribution 
are based on the work in Ref.~\cite{Abdulhamid:2021xtt}. 
Many thanks to all co-authors for collaboration. 
We are grateful to the organizers of DIS2022 for the 
invitation to present these results at the workshop.

\nolinenumbers

\end{document}